\documentclass{PoS}

\usepackage{enumerate}

\def\eg{{e.g.,~}}
\def\ie{{i.e.,~}}

\title{Extragalactic background light inferred from AEGIS galaxy SED-type fractions}

\ShortTitle{EBL inferred from AEGIS galaxy SED-type fractions}

\author{\speaker{A. Dom\'inguez}\thanks{Visiting Student at the Santa Cruz Institute for Particle Physics (SCIPP), University of California, Santa Cruz.}\\
Universidad de Sevilla \& IAA-CSIC\\
E-mail: \email{alberto@iaa.es}}

\author{J.~R. Primack,$^a$ D.~J. Rosario,$^a$, F. Prada,$^b$ R.~C. Gilmore,$^{ac}$ S.~M. Faber,$^a$ D.~C. Koo,$^a$ R.~S. Somerville,$^d$ M.~A. P\'erez-Torres,$^b$ P. P\'erez-Gonz\'alez,$^e$ J.-S. Huang,$^f$ M. Davis,$^g$ P. Guhathakurta,$^a$ P. Barmby,$^h$, C. J. Conselice,$^i$ M. Lozano$^j$, J.~A. Newman$^k$ and M.~C. Cooper$^l$\\
\llap{$^a$}University of California Santa Cruz;~
\llap{$^b$}IAA-CSIC;~
\llap{$^c$}Scuola Internazionale Superiore di Studi Avanzati;~
\llap{$^d$}Space Telescope Science Institute;~
\llap{$^e$}Universidad Complutense de Madrid;~
\llap{$^f$}Harvard-Smithsonian Center for Astrophysics;~
\llap{$^g$}University of California Berkeley;~
\llap{$^h$}University of Western Ontario;~
\llap{$^i$}University of Nottingham;~
\llap{$^j$}Universidad de Sevilla;~
\llap{$^k$}University of Pittsburgh;~
\llap{$^l$}University of Arizona
}

\abstract{The extragalactic background light (EBL) is of fundamental importance both for understanding the entire process of galaxy evolution and for $\gamma$-ray astronomy. However, the overall spectrum of the EBL between 0.1 and 1000~$\mu$m has never been determined directly neither from observed luminosity functions (LFs), over a wide redshift range, nor from any multiwavelength observation of galaxy spectral energy distributions (SEDs). The evolving, overall spectrum of the EBL is derived here utilizing a novel method based on observations only. The changing fractions of quiescent galaxies, star-forming galaxies, starburst galaxies and active galactic nucleus (AGN) galaxies from redshift 0.2 to 1 are estimated, and two alternative extrapolations of SED types to higher redshifts are considered. This allows calculation of the evolving EBL. The EBL uncertainties in our modelling based directly on the data are quantified, and their consequences for attenuation of very-high-energy $\gamma$-rays due to pair production on the EBL are discussed. It is concluded that the EBL seems well constrained from the UV to the mid-IR at an intensity level roughly matching galaxy count data. Independent efforts from IR and $\gamma$-ray astronomy are needed in order to reduce the uncertainties in the far-IR.}

\FullConference{25th Texas Symposium on Relativistic Astrophysics \\
		 December 6-10, 2010\\
		 Heidelberg, Germany}

\begin{document}

\section{Introduction}
\label{sec:intro}
The formation and evolution of galaxies in the universe are accompanied unavoidably by the emission of radiation. All this radiated energy is still streaming through the universe, although much is now at longer wavelengths due to redshifting and absorption/re-emission by dust. The photons mostly lie in the range of $\sim$0.1-1000~$\mu$m, \ie ultraviolet (UV), optical and infrared (IR), and produce the second-most energetic diffuse background after the cosmic microwave background, thus being essential for understanding the full energy balance of the universe. We will account in this work for the radiation accumulated by star formation processes through most of the life of the universe, plus a contribution from active galactic nuclei (AGNs) to this wavelength range, known as the diffuse extragalactic background light (EBL). The direct measurement of the EBL is a very difficult task subject to high uncertainties. This is mainly due to the contribution of zodiacal light, some orders of magnitude larger than the EBL (\eg \cite{hauser01,chary10}). Other observational approaches set reliable lower limits on the EBL, such as measuring the integrated light from discrete extragalactic sources (\eg \cite{madau00,fazio04,keenan10}). On the other hand, there are phenomenological approaches in the literature that predict an overall EBL model (\ie between 0.1 and 1000~$\mu$m and for any redshift). These are basically of the following four kinds.

\begin{enumerate}[(i)]
\item Forward evolution, which begins with cosmological initial conditions and follows a forward evolution with time by means of semi-analytical models (SAMs) of galaxy formation (\eg \cite{somerville11,gilmore11}. These two papers describe the same modelling but discussed in different contexts).

\item Backward evolution, which begins with existing galaxy populations and extrapolates them backwards in time (\eg \cite{stecker06,franceschini08}).

\item Evolution of the galaxy populations that is inferred over a range of redshifts. The galaxy evolution is inferred here using some quantity derived from observations such as the star formation rate (SFR) density of the universe (\eg \cite{finke10,kneiske10}).

\item Evolution of the galaxy populations that is directly observed over the range of redshifts that contribute significantly to the EBL. The methodology presented in Ref.~\cite{dominguez11}, which we term empirical, is the first that belongs to this category. The main results presented in that paper are summarized here.

\end{enumerate}

Our EBL estimates from type (iv) methodology are compared with the results from the approach type (i) presented in Ref.~\cite{somerville11,gilmore11}, and with the results from type (ii) methodology by Ref.~\cite{franceschini08}. We consider these two models the most robust among the literature. First, we consider the theoretical approach taken in Refs.~\cite{somerville11,gilmore11} as complementary to our observationally motivated one to eventually reach a complete understanding of galaxy evolution. Second, approaches type (ii) are potentially problematic because they imply extrapolations backwards in time of local or low-redshift luminosity functions (LFs). Intrinsically different galaxy populations exist at high redshifts, which cannot be accounted for by these extrapolations. Particularly, \cite{franceschini08} use observed LFs in the near-IR from the local universe to $z=1.4$ for describing the elliptical and spiral populations, and only local for describing irregular/starbursting galaxies. They distinguish between these galaxy morphologies using images from different satellites. Different local LFs and data sets in the IR are used to constrain the mid and far-IR background. Their modelling is complex and not reproducible. Despite these particular problems, this methodology is based upon LFs, quantity directly observed and well understood unlike type (iii) models based on parametrizations of the history of the SFR density of the universe, quantity with large uncertainties and biases.

One important application of the EBL for $\gamma$-ray astronomy is to recover the unattenuated spectra of extragalactic sources. Our goal is to measure the EBL with enough precision that the uncertainties due to the EBL modelling, in these recovered unattenuated spectra, are small compared with other effects such as statistical or systematic uncertainties in the $\gamma$-ray observations. Examples of this are discussed in Sec.~\ref{sec:atte}.

Throughout this work, a standard $\Lambda$CDM cosmology is assumed, with matter density $\Omega_{m}=0.3$, vacuum energy density $\Omega_{\Lambda}=0.7$, and Hubble constant $H_{0}=70$~km~s$^{-1}$Mpc$^{-1}$.

\section{Methodology}
\label{sec:method}
Our model is based on the rest-frame $K$-band galaxy LF in Ref.~\cite{cirasuolo10} and on multiwavelength galaxy data from the All-wavelength Extended Groth Strip International Survey (AEGIS\footnote{http://aegis.ucolick.org/}, \cite{davis07}) of about 6000 galaxies in the redshift range of 0.2-1. These data sets are put together in a very transparent and consistent framework. The \cite{cirasuolo10} LF is used to count galaxies (and therefore to normalize the total EBL spectral intensity) at each redshift. The LF as well as our galaxy sample is divided into three magnitude bins according to the absolute rest-frame $K$-band magnitude, \ie faint, middle and bright. Within every magnitude bin, an SED type is statistically attached to each galaxy in the LF assuming SED-type fractions that are a function of redshift within those magnitude bins. This is estimated by fitting our AEGIS galaxy sample to the 25 galaxy-SED templates from the SWIRE\footnote{http://www.iasf-milano.inaf.it/$\sim$polletta/templates/swire$\_$templates.html} library \cite{polletta07}. Then, luminosity densities are calculated from these magnitude bins from every galaxy population at all wavelengths, and finally all the light at all redshifts is added up to get the overall EBL spectrum. In this approach, it is possible to directly calculate the contribution to the EBL from all redshift bins, as well as the evolution of the EBL spectrum with redshift and the processes related to this evolution, by sources of all the 25 SED types considered.

A multiwavelength galaxy catalogue built from AEGIS for this work is used. This catalogue contains 5986 galaxies, all in the Extended Groth Strip (EGS). It is required that every galaxy in the sample have 5$\sigma$ detections in the $B$, $R$, $I$, $K_{S}$ and Infrared Array Camera (IRAC) 1 bands, and observations (but not necessarily detections) in the IRAC~2, 3, 4 and Multiband Imaging Photometer for Spitzer (MIPS) 24 bands. These 5$\sigma$ upper limits are given by the following fluxes: 1.2, 6.3, 6.9 and 30~$\mu$Jy for IRAC~2, 3, 4 and MIPS~24, respectively, according to Ref.~\cite{barmby08} for the IRAC bands and
Ref.~\cite{dickinson07} for MIPS~24. In addition, 1129 of these galaxies have Galaxy Evolution Explorer (GALEX) detections in the far-UV and 2345 galaxies in the near-UV. In our sample, 4376 galaxies have the highest quality spectroscopic redshifts measured by the Deep Evolutionary Exploratory Probe 2 team (DEEP2 DR3, \cite{newman11}), with the Deep Imaging Multi-Object Spectrograph (DEIMOS) spectrograph \cite{faber03} on the Keck II telescope in an area of about 0.7~deg$^{2}$ in the sky. All the other galaxies in the sample (1610 galaxies) have secure photometric redshifts, more than 80 per cent with uncertainty in redshift less than 0.1.

\section{Results}
\label{sec:results}
\subsection{Galaxy SED-type fractions}
The \texttt{Le PHARE} code is used to fit every galaxy in our sample to the 25 SWIRE templates. For clarity, we will compress in our discussion (but not in our calculations, where they will remain independent) the 25 SED types in the SWIRE library to four groups: quiescent galaxies, star-forming galaxies, starbursts and AGN galaxies. 

To avoid accounting for bad fits, which do not correctly describe the galaxy photometric data, a cut in $\chi^{2}_{red}=\chi^{2}/n$ is applied, with $\chi^{2}$ given by \texttt{Le PHARE} and $n$ degrees of freedom (bands with detections). We have checked carefully that $\chi^{2}_{red}\le 30$ is a red good value for quiescent, star-forming and starburst galaxies, but AGN galaxies are systematically worse fits, probably due to the fact that there is a large range in AGN SED shapes due to multiple emission components which cannot be easily encapsulated in a few templates and a stronger cut is needed $\chi^{2}_{red}\le 20$.

\begin{figure}
\centering
\includegraphics[width=8cm]{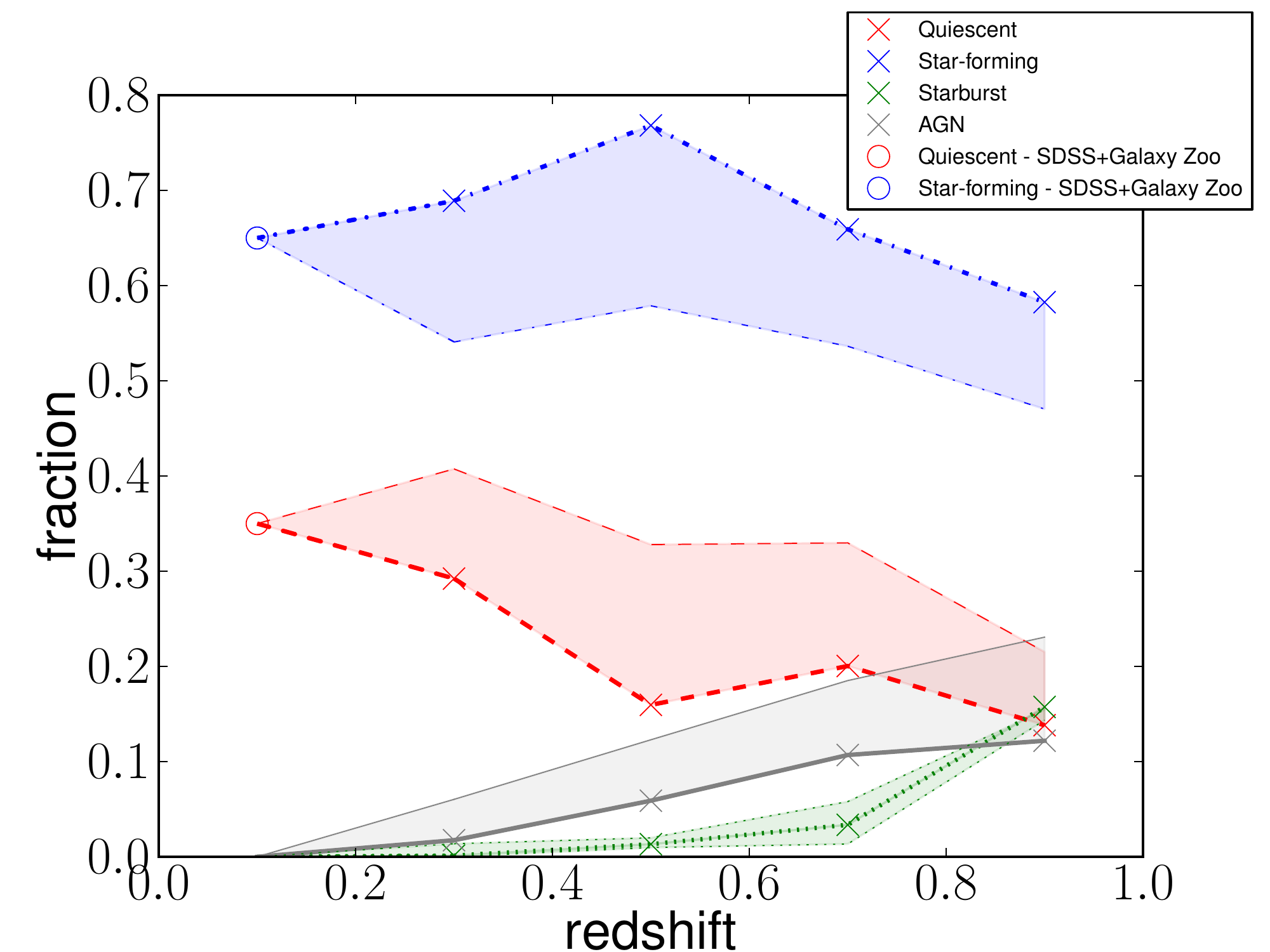}
\caption{Galaxy SED-type fractions from our catalogue (after the $\chi_{red}^{2}$ cuts) of the different populations versus redshift according to our multiwavelength fits. We mark with crosses our fractions from $z=0.9-0.3$. The circles at $z=0.1$ are fractions computed from the SDSS-based sample (see Ref.~\cite{dominguez11} for details). We show with a shadow area the uncertainties from our lower limit for the errors as well as for our $\chi^{2}_{red}$ cut for fits. The uncertainties are around $\pm 0.1$.}
\label{fig:frac}
\end{figure}

Fig.~\ref{fig:frac} shows the galaxy-SED-type fractions for four different redshift bins up to $z = 1$, where we have chosen bins of $z = 0.2$ for statistical reasons. The redshift range shown corresponds almost to 60 per cent of the age of the universe. The shadow regions are the uncertainties due to the lower limits on the photometric errors for the catalogue and for the $\chi^{2}_{red}$ cuts. This region is calculated changing the lower limits from 1 to 10 per cent in steps of 1 per cent and applying extreme cases for the cuts for every lower limit. The boundaries from these calculations lead to the shadow regions. The fractions adopted for the model are marked with crosses and wider lines. We observe that the fraction of quiescent galaxies (dashed-red line) increases by a factor of $\sim 2$ from $z\sim 0.9$ to 0. 3, while the star-forming fraction (dotted-dashed-blue line) keeps roughly constant for the full redshift range peaking at $z = 0.5$. Starburst-type galaxies (dotted-green line) decrease very quickly from $z\sim 0.9$ and reach almost 0 at $z\sim 0.5$. On the other hand, the AGN-type fraction (solid-gray line) is roughly constant from $z\sim 0.9$ to 0.7 and then decreases to 0.02 at $z\sim 0.3$. This result should not be considered a complete picture of the evolution of the galaxy populations in the universe since these fractions depend on the color-magnitude limits of the survey. But what is certainly described is the population of galaxies that contribute the most to the EBL around the knee of the LF.

For the high-redshift universe ($z > 1$, where there are no galaxies in our sample), two different cases are considered for the evolution of the galaxy-SED-type fraction. It is shown that our results are not changed significantly except in the far-IR by these two choices. For the redshifts less than those of the most distant known $\gamma$-ray sources, and redshifts where sources are likely to be found in the near future by Imaging Atmospheric Cherenkov Telescopes (IACTs), we find that there is almost no change in the EBL even with a fairly large adjustment in the evolution of galaxy-SED-type fractions. The fiducial choice is to keep constant the fractions computed for our highest redshift bin. This choice is made for simplicity, due to the difficulty in the multiwavelength classification of distant galaxies with current instruments. As an alternative approach, we choose to increase linearly with redshift the starburst-like fraction from our calculated 16 per cent at $z = 0.9$ up to 60 per cent at $z = 2$, while decreasing at the same rate the quiescent and star-forming galaxies. The weight of each of the 25 SWIRE templates is changed in the same proportion. The fractions are kept constant at $z = 2$ for $z > 2$. This approach is called high starburst and it is used to determine a likely upper limit on the EBL at long wavelengths

\subsection{Extragalactic background light}
\begin{figure}
\centering
\includegraphics[width=10cm]{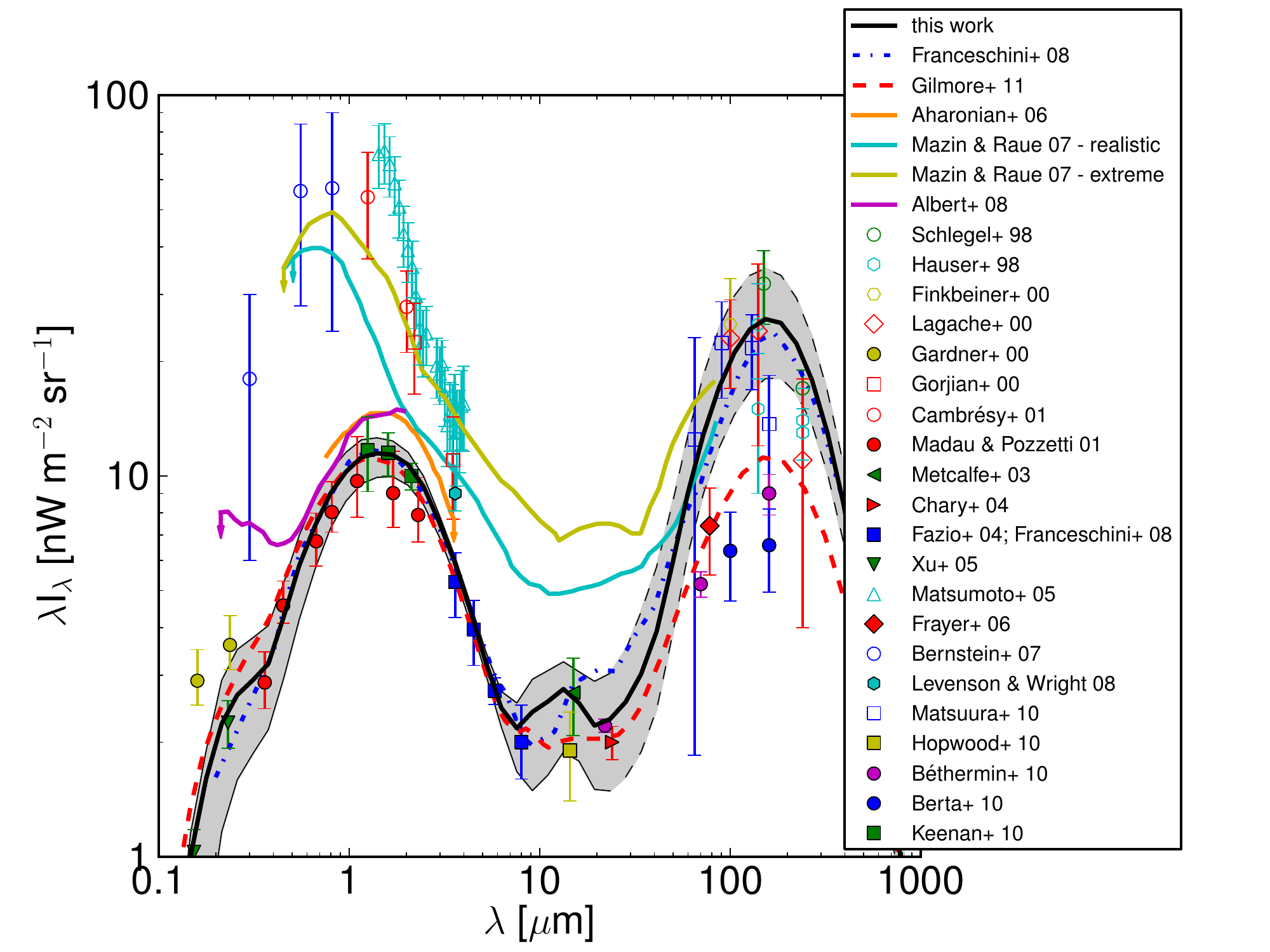}
\caption{The solid-black line is the extragalactic background light calculated by the fiducial extrapolation of the galaxy SED-type fractions for $z > 1$. Uncertainties in the our EBL estimation are shown with a shadow area (see Ref.~\cite{dominguez11} for a discussion on this and for details on the references). The envelope of the shadow region within the dashed line at wavelengths above 24~$\mu$m shows the region where there is no photometry in our galaxy catalogue.}
\label{fig:ebl}
\end{figure}

The local galaxy luminosity density, its evolution over redshift at different wavelengths for the two extrapolations previously considered for the high-redshift fractions, as well as an estimation of the SFR density of the universe are studied in Ref.~\cite{dominguez11}. Other quantities such as the EBL evolution\footnote{EBL specific intensities are publicly available at http://side.iaa.es/EBL} are discussed there as well. Here, we only show in Fig.~\ref{fig:ebl} the EBL in the local universe, with its uncertainties, compared with direct and indirect observational data, and other EBL models. Fig.~\ref{fig:ebl} suggests that the EBL coming from galaxies is already well constrained in the region from the UV up to the mid-IR but not in the far-IR. Galaxy counts from very deep surveys taken with very sensitive instruments \cite{madau00,fazio04,keenan10} should be considered as a good estimation of the true EBL from galaxies. On the other hand, different fully independent modelings based on different galaxy data sets \cite{gilmore11,franceschini08,dominguez11} agree in the specific intensity level of the EBL. In particular, galaxy count data are in excellent agreement with our results.

\section{Application to $\gamma$-ray attenuation}
\label{sec:atte}
The EBL has important implications for the interpretation of data taken using recent $\gamma$-ray experiments such as the Fermi satellite and IACTs (VERITAS, HESS, and MAGIC) due to the photon-photon pair production between $\gamma$-ray photons traveling across cosmological distances and EBL photons \cite{nikishov62,gould66}. Blazars are an important source of extragalactic $\gamma$-ray emission and have become a relevant tool for indirectly measuring the EBL. These objects are believed to be an extreme category of AGNs. Their emission, which occurs at all wavelengths of the electromagnetic spectrum, comes from super-massive black holes (with masses of $\ge 10^{7}$~M$_{\odot}$) swallowing matter accreted from their surroundings. In general, AGNs are characterized by a beamed emission perpendicular to the accretion disc known as jets, which are pointing towards us in the case of blazars.

\begin{figure}
\includegraphics[width=\columnwidth]{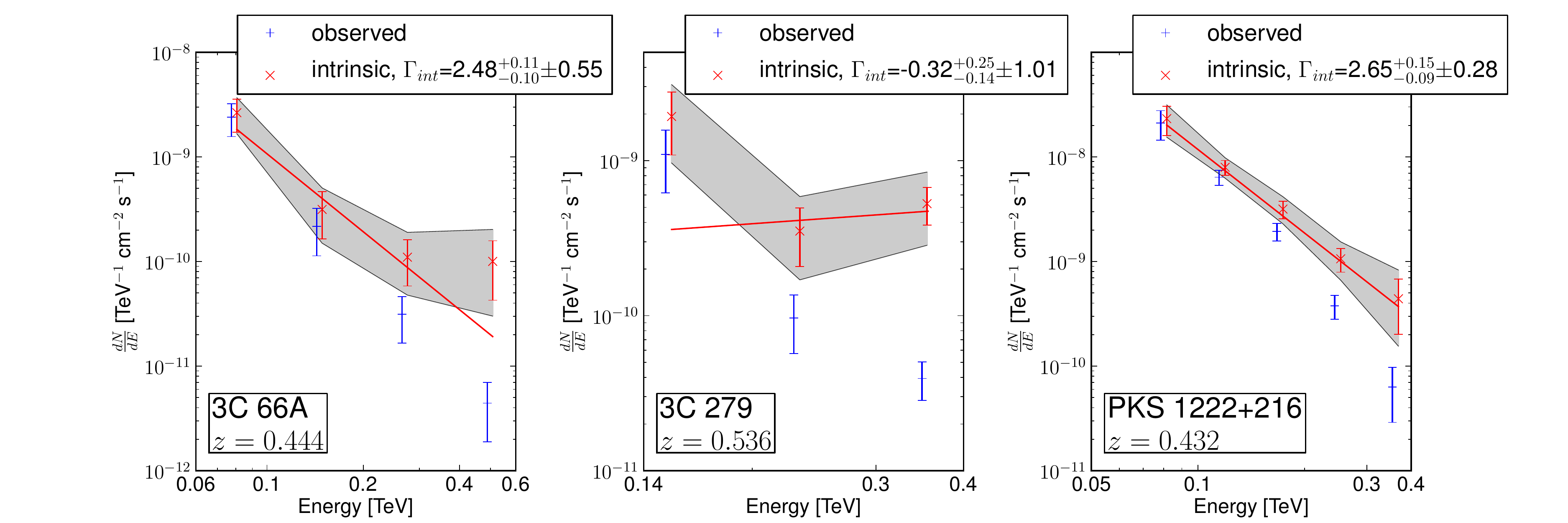}
\caption{VHE spectra observed (blue) and EBL corrected (red) from the attenuation calculated with our EBL model of three high-redshift blazars measured by MAGIC \cite{aleksic11a,aleksic11b,aleksic11c}. The straight-red line is the best-fitting power law for every blazar with index $\Gamma_{int}$. Uncertainties from the EBL modelling as well as statistical and systematic errors are shown with a shadow region. The first uncertainties in the index are due to the EBL modelling and the second uncertainties are statistical plus systematic errors. Intrinsic energies are slightly shifted for clarity.}
\label{fig:blazars}
\end{figure}

Following both theoretical arguments \cite{boettcher07,sikora09} and observational facts \cite{hartman99,abdo10}, it is assumed that no intrinsic (or EBL-corrected) VHE spectra from blazars might be fitted to a power-law with indexes harder than 1.5. We now proceed to test in Fig.~\ref{fig:blazars} whether the observed spectra of three new measurements of high-redshift AGNs (other different spectra were considered in Ref.~\cite{dominguez11}) satisfy the condition that the intrinsic spectrum corrected by the attenuation derived with our EBL model\footnote{Optical depths are publicly available at http://side.iaa.es/EBL} has $\Gamma_{int}\ge 1.5$. We consider the following blazars: 3C~66A at $z = 0.444$ observed by MAGIC \cite{aleksic11a}, 3C~279 at $z=0.536$ observed by MAGIC in the 2007 observational campaign \cite{aleksic11b}, and the discovery of PKS~1222+216 in the VHE regime \cite{aleksic11c}, the second most distant flat-spectrum radio quasar known ($z=0.432$). These three blazars are plotted in Fig.~\ref{fig:blazars}, where the legends show that the condition $\Gamma_{int}\ge 1.5$ is satisfied. We note that in the 3C~279 case, only having three data points makes the fit no statistically reliable.

It is confirmed from the study of these blazars the conclusions obtained in Ref.~\cite{dominguez11}. First, our EBL is generally compatible with the expected hardness of the EBL-corrected slopes. However, it is clear that a simple SSC model cannot explain any flatness at the highest energies of the EBL-corrected spectra of 3C~66A, which suggests that some extension to the model may be necessary such as an external photon region, a better understanding of the IACT systematic uncertainties or even a revision of the propagation mechanisms mainly through the intergalactic medium \cite{sanchez-conde09}.

Second, the uncertainties in the EBL-corrected spectra are dominated by other effects different than EBL modelling as shown in the index uncertainties in Fig.~\ref{fig:blazars}.

\section{Conclusions}
\label{conclusions}
A novel, robust, and powerful method based on observations to derive the evolving spectrum of the EBL between 0.1 and 1000~$\mu$m was thoroughly presented in Ref.~\cite{dominguez11} and reviewed here. This model is based on the observed rest-frame $K$-band galaxy LF over redshift found in Ref.~\cite{cirasuolo10}, combined with an estimation of galaxy-SED-type fractions based on a multiwavelength sample of $\sim 6000$ galaxies from AEGIS. This model has the following main advantages over other existing EBL models: transparent methodology, reproducibility, and utilizing direct galaxy data. The best available data sets are used (\cite{cirasuolo10}'s LF and the AEGIS galaxy catalogue) observed over a wide redshift range. The galaxy evolution is directly observed in the rest-frame $K$ band up to $z = 4$. Observed galaxies up to $z = 1$ from the UV up to 24~$\mu$m with SEDs of 25 different types (from quiescent to rapidly star-forming galaxies and including AGN galaxies) are taken into account in the same observational framework. A study of the uncertainties to the model directly from the data (such as uncertainties in the Schechter parameters of the \cite{cirasuolo10} LF and the errors in the photometric catalogue) is done, and their propagated uncertainties to the $\gamma$-ray attenuation are studied. Two extrapolations of the galaxy-SED-type fractions to $z > 1$ were considered, showing that these assumptions only affect the far-IR, where the uncertainties of the modelling are the largest because of the lack in our catalogue of far-IR photometry and the poor understanding on galaxy SED at large redshifts.

We concluded that the EBL from galaxies seems already well constrained from UV to mid-IR wavelengths, even though uncertainties are still large in the far-IR. Furthermore, discoveries of $\gamma$-ray from distant blazars (\eg \cite{aleksic11a,aleksic11b,aleksic11c}) support the EBL specific intensity level derived from galaxy count and recent EBL models \cite{gilmore11,franceschini08,dominguez11}. As discussed the VHE recovered spectra of blazars up to $\sim 10$~TeV are dominated by statistical and systematic uncertainties in the observations rather than EBL modelling. We highlight that the EBL specific intensity calculated with our method is matching the lower limits from galaxy counts, which implies the highest transparency of the universe to $\gamma$-ray allowed by standard physics. This predicts a promising future for the new generation of IACTs, namely CTA.

\section*{Acknowledgements}
A. Dom\'inguez thanks the financial support of a Fermi grant to participate in the 25th Texas Symposium on Relativistic Astrophysics and to M.~A. S\'anchez-Conde for helpful comments.

\end{document}